\documentclass{aa}  

\usepackage{graphicx}
\usepackage{txfonts}
\begin{document}

\title{AA Tau's sudden and long-lasting deepening: enhanced extinction
  by its circumstellar disk\thanks{Based on observations made with
    telescopes at the Crimean Astrophysical Observatory, at the Calar
    Alto Observatory, and at ESO Paranal Observatory under programme
    ID 288.C-5026(A).}}

   \author{J. Bouvier
          \inst{1}
          \and
          K. Grankin\inst{2}
          \and
          L.~E. Ellerbroek\inst{3}
          \and
          H. Bouy\inst{4}
          \and
          D. Barrado\inst{4}
         }

   \institute{UJF-Grenoble 1 / CNRS-INSU, Institut de Planétologie et d'Astrophysique de Grenoble (IPAG) UMR 5274, Grenoble, F-38041, France\\
              \email{Jerome.Bouvier@obs.ujf-grenoble.fr}
         \and
            Crimean Astrophysical Observatory, Scientific Research
            institute, 98409, Ukraine, Crimea, Nauchny
             \and 
             Astronomical Institute Anton Pannekoek, Science Park 904,
             P.O. Box 94249, 1090 GE Amsterdam, The Netherlands 
             \and
             Centro de Astrobiologa (INTA-CSIC); PO Box 78, 28691,
             Villanueva de la Cañada, Spain
            }

   \date{Received March 1, 2013; accepted March 25, 2013}

% \abstract{}{}{}{}{} 
% 5 {} token are mandatory
 
  \abstract
  {AA Tau has been monitored for more than 20 years since 1987 and
    exhibited a nearly constant brightness level of V=12.5~mag. We
    report here that in 2011 it suddenly faded away, becoming 2
    magnitudes fainter in the V-band, and has remained in this deep
    state since then.}
   {We investigate the origin of the sudden dimming of the AA Tau system.}
   {We report on new optical and near-IR photometry and spectroscopy obtained
     during the fading event.}
   {The system appears to be much redder and fainter than it was in the
     bright state. Moreover, the 8.2d photometric period continuously
     observed for more than 20~years is not detected during most of the
     deep state. The analysis of the system's brightness and colors
     suggests that the visual extinction on the line of sight has
     increased by about 3-4 magnitudes in the deep state. At optical
     wavelengths, the system appears to be dominated by scattered
     light, probably originating from the upper surface layers of a
     highly inclined circumstellar disk. The profiles of the Balmer
     lines have significantly changed as well, with the disappearance
     of a central absorption component regularly observed in the
     bright state. We ascribe this change to the scattering of the
     system's spectrum by circumstellar dust. Remarkably, the mass
     accretion rate in the inner disk and onto the central star has
     not changed as the system faded. }
   {We conclude that the deepening of the AA Tau system is due to a
     sudden increase of circumstellar dust extinction on the line of
     sight without concomitant change in the accretion rate. We
     suggest that the enhanced obscuration may be produced by a
     non-axisymmetric overdense region in the disk, located at a
     distance of 7.7~AU or more, that was recently brought into the line
     of sight by its Keplerian motion around the central star. }


   \maketitle

\section{Introduction}

%cf. W.-P. Chen et al. 2012 for a similar deepening ; also Grinin et al. 1994
%for UX Ori type variability

A young, low-mass member of the 1-3~Myr Taurus association (Kenyon \&
Hartmann 1995), AA~Tau has provided many clues to the
magnetically mediated accretion process occurring in young stellar
objects. The strong large-scale magnetic fields measured at the
surface of young low-mass stars (Donati \& Landstreet 2009) are
thought to be able to disrupt the inner accretion disk surrounding the
central star up to a distance of several stellar radii (Camenzind
1990; K\"onigl 1991). From the inner disk edge to the stellar surface,
magnetically controlled accretion funnel flows develop, eventually
reaching the stellar surface at nearly free-fall velocity, thus
creating a localized accretion shock seen as a hot spot (e.g., Bertout
et al. 1988). In addition, the magnetic star-disk interaction is
expected to create a nonaxisymmetric disturbance at the inner disk
edge, i.e., an inner disk warp, in the case of an inclined
magnetosphere (e.g. Terquem \& Papaloizou 2000). Many clues have been
gathered that suggest that magnetospheric accretion is quite a common
process indeed in young accreting low-mass stars (see, e.g., Bouvier
et al. 2007a for a review).

Thanks to its close to edge-on orientation on the line of sight
($i\simeq$70-75$\degr$, Bouvier et al. 2003; O'Sullivan et al. 2005; Cox
et al. 2013), AA~Tau has yielded deep insight into the actual geometry
and dynamics of the magnetospheric accretion process.
Spectrophotometric monitoring campaigns performed on this source
(cf. Bouvier et al. 1999, 2003, 2007) and polarimetric studies
(M\'enard et al. 2003; O'Sullivan et al. 2005) have investigated the
geometry of the magnetically controlled accretion flows from the inner
disk to the stellar surface, as well as the connection between the
funnel flows and the hot spots at the stellar surface. They have
also revealed a nonaxisymmetric inner disk warp,
presumably driven by the interaction of the disk with an inclined
dipole, and have highlighted the very dynamical nature of this
interaction, with significant structural changes occurring on a
timescale of weeks. Direct magnetic field measurements have confirmed
 an inclined 2-3~kG magnetic dipole at the surface of
AA Tau (Donati et al. 2010), and the circumstellar disk has been
recently directly imaged, confirming its warped structure (Cox et
al. 2013).

While AA Tau has served as a Rosetta Stone for the understanding of
the geometry and dynamics of the accretion flow in young magnetic
stars, many other cases have now been reported that display similar
photometric variations and associated spectroscopic signatures.
Indeed, a distinctive signature of AA Tau-like light curves is the
occurrence of quasi-periodic dips that interrupt a nearly constant
bright photometric level. The dips occur on a week's timescale and are
interpreted as resulting from the partial occultation of the central
star by the warped inner disk edge as it rotates at Keplerian velocity
at a distance of a few stellar radii from the stellar surface (Bouvier
et al. 1999). In a large-scale Corot-based photometric study of the
3~Myr NGC~2264 star-forming region, Alencar et al. (2010) reported a
fraction of 40$\pm$10\% of AA Tau-like light curves among classical T
Tauri stars with optically thick accretion disks. Hence, the AA Tau
phenomenon, i.e., the quasi-periodic obscuration of the central star
by a nonaxisymmetric inner disk, is found to be fairly common, as
long as a proper sampling of the photometric variations is secured (see
also Artemenko et al. 2012; Cody et al. 2013).

AA Tau itself has remained at a nearly constant maximum brightness
level (V$\simeq$12.4-12.6~mag) between 1987 and 2010, i.e., for at
least 24 years (Grankin et al. 2007).  We report here that it suddenly
lost about 2~magnitudes in 2011 and has remained in this deep state
until now (March~2013). We present here photometric and spectroscopic
observations obtained during the deep state of the system.  In
Section~\ref{obs}, we detail the photometric and spectroscopic
observations. In Section~\ref{res}, we present the photometric and
spectroscopic properties of the system in the deep state. In
Section~\ref{dis}, we show that the current state of the AA Tau system
can be consistently interpreted as resulting from a sudden increase of
extinction on the line of sight and discuss its possible origin in the
circumstellar disk.

\section{Observations}
\label{obs}

\subsection{Photometry}

Optical observations of AA Tau were obtained at the Crimean
Astrophysical Observatory (CrAO) from October 2007 to February
2013. The measurements were carried out with the CrAO 1.25m telescope
(AZT-11) equipped with either the Finnish five-channel photometer or
the FLI~1001E CCD detector. The photomultiplier tube was used during
the bright phases of the star (V$\leq$14 mag) and the CCD detector for
the fainter phases. All photometric observations were obtained in the
V and R$_J$ filters. Differential photometry was performed on CCD
images and absolute photometry from photomultiplier observations, with
an accuracy of about 0.05~mag in both filters. The two
comparison stars are located less than 4~arcmin away from AA Tau
(these are J04344511+2424432 and J04344826+2428532 in the 2MASS
catalog). All data reduction procedures followed those previously
described in Bouvier et al. (2003).

Additional BVR$_c$I$_c$ photometry was obtained on December 23, 2011
using the Cafos focal reducer (Meisenheimer 1994) in direct-imaging
mode with CCD SITE1d\_15 on the 2.2m Calar Alto Telescope. Individual
BVRI images were bias subtracted and flat-field
corrected. Differential photometry was performed between AA Tau and the
same comparison stars as at CrAO. The resulting photometry is accurate to
about 0.05~mag.

Near-infrared JHK-photometry was obtained on December 13, 17, and 18, 2011
with the Omega2000 camera (Kovacs et al. 2004) mounted on the 3.5m
Calar Alto telescope. The O2000 individual images were first processed
using the Alambic EIS-MVM pipeline (Vandame 2002) adapted to the
characteristics of the instrument. Briefly, the pipeline includes
standard pre-processing procedures (overscan and bias subtraction,
flat-fielding), followed by an unbiased two-pass background subtraction,
registration, and stacking of the individual images. A second-order
astrometric solution was derived using the 2MASS catalog as
reference. Differential photometry was obtained with the same neighboring
comparison stars, whose JHK photometry was obtained from the 2Mass
catalog. 

The full set of photometric measurements obtained at CrAO from 2007 to
2013 is available in electronic form. The additional photometry
obtained at Calar Alto Observatory is listed in Table~\ref{phot}. 

\begin{table}
\caption{CAHA optical and near-IR photometry of the system.}             % title of Table
\label{phot}      % is used to refer this table in the text
\centering                          % used for centering table
\begin{tabular}{l l l l l }        % centered columns (4 columns)
\hline\hline                 % inserts double horizontal lines
%\multicolumn{5}{c}{Calar Alto  Cafos  optical photometry } \\
JD2400000   & V   & B &  Rc   &     Ic\\
55919.5426  & 15.578  & 17.145 & 14.351  & 12.943 \\
\hline                        % inserts single horizontal line
%\multicolumn{4}{c}{Calar Alto  Omega 2000  JHK photometry} \\
JD2400000      &J      &H      &K \\
55909.3290      &10.464     & 9.202    &  8.458 \\
55913.4358      &10.614     & 9.266     & 8.431 \\
55914.6103      &10.525     & 9.210     & 8.446 \\
\hline                                   %inserts single line
\end{tabular}
\end{table}

%
%2MASS 50782.8175       9.433       8.546      8.047

\subsection{Spectroscopy}

Low-resolution ($\sim$0.7~nm) optical spectroscopy was performed on 13
and 23 December 2011 using the Cafos focal reducer (Meisenheimer 1994) on
the 2.2m Calar Alto telescope equipped with the g-100 grism. The
exposure time was 1800s for each spectrum, and the slit width was 
$1\farcs5$. Two 1800s spectra were obtained in sequence on Dec.13,
2011, and were subsequently averaged.The spectrophotometric standard
G191B2b was observed at the same airmass as AA Tau on both nights.
After the 2D images were bias corrected and flat-fielded, the object's
and neighboring sky's spectra were extracted using the IRAF/twodspec
package. The wavelength was calibrated using the HgHeRb spectral
lamp. Owing to the low number of lines available in the arc spectrum
over the 480-770~nm range, the wavelength calibration is not accurate
to better than 2~nm.  Finally, the response of the instrument was
corrected for using the spectrum of the spectrophotometric standard
G191B2b.

Medium-resolution spectroscopy was obtained with ESO VLT/XSHOOTER on
25 February 2012. XSHOOTER is a cross-dispersed echelle spectrograph
with three arms: UVB (300--590 nm), VIS (550--1020~nm) and NIR
(1000--2480~nm) (D'odorico et al. 2006; Vernet et al. 2011). Slit
widths of $1\farcs3$, $0\farcs7$ and $0\farcs4$ were chosen in these
arms to deliver a corresponding spectral resolution of 4000, 11000,
and 11300 in the respective wavelength domains. The standard steps of
data reduction (order extraction, flat-fielding, wavelength
calibration) were applied using the XSHOOTER pipeline (Modigliani et
al. 2010). The observations comprised four exposures at alternating
offsets along the slit (nodding), allowing for accurate sky
subtraction, with a total exposure time of 1000~sec. The resulting
one-dimensional spectra were flux-calibrated using a spectrum of the
spectrophotometric standard star LTT3218, which was taken directly
before AA Tau. We corrected for slitlosses, taking into account the
average seeing ($1\farcs 0$ in V), resulting in a scaling up of 1.24,
1.93 and 3.21 in the UVB, VIS and NIR arms, respectively. The
resulting flux calibration agrees well in the overlapping wavelength
regions between the arms.

% XSH spectrum actually on 26.02.2012UT : 2012-02-26T00:33:47.888 = JD55983.02347093

\section{Results}
\label{res}

\subsection{Photometric variability}

\begin{figure*}
   \centering
  \includegraphics{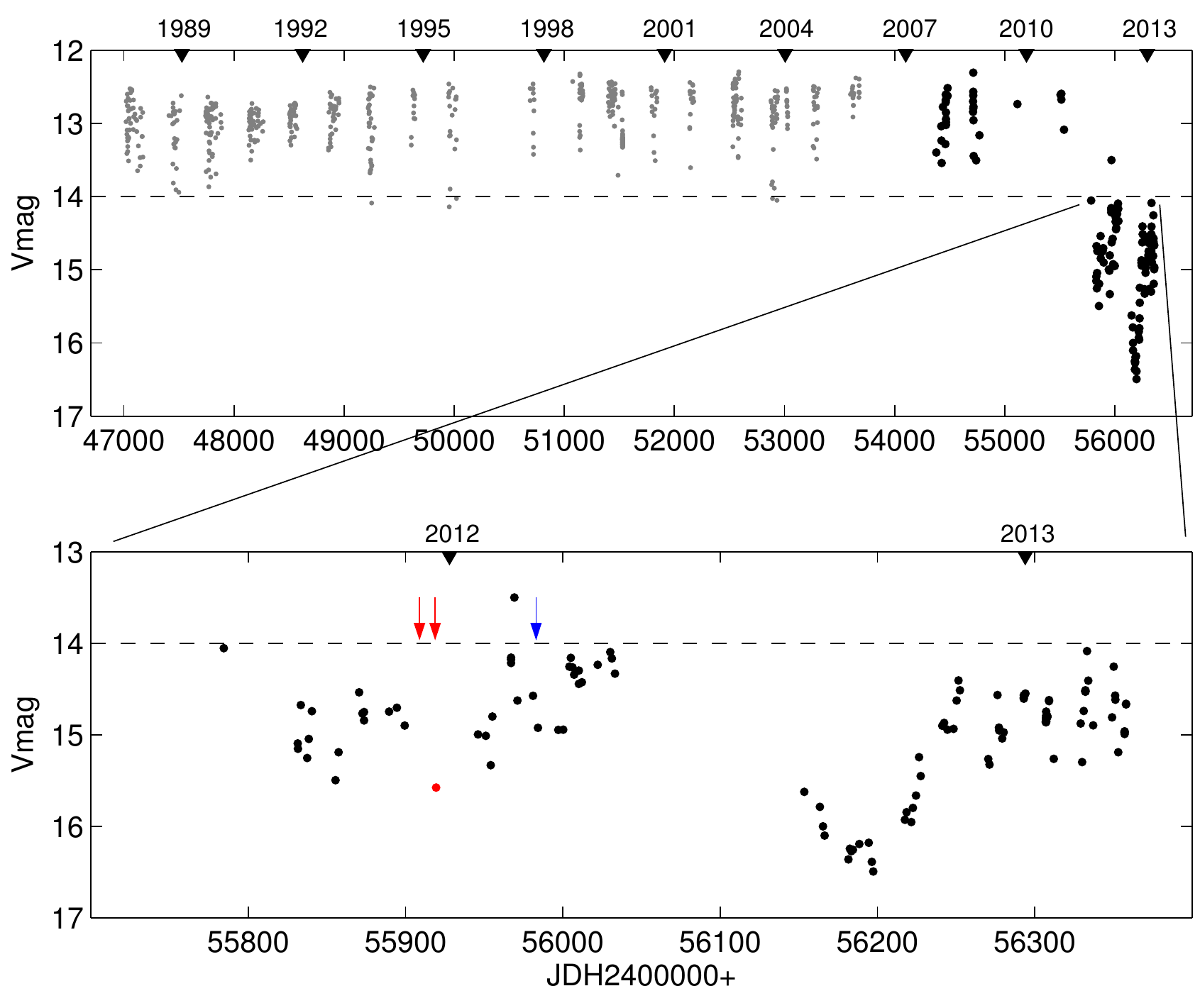}
  \caption{V-band AA Tau light-curve spanning 26 years is shown on
    the top panel. Gray circles refer to Mt Maidanak measurements
    (Grankin et al. 2007) while dark circles are new measurements from
    CrAO. The V=14~mag brightness level is shown as
    dashed line for reference. The bottom panel shows a zoom on the two last observing
    seasons (2011-2012). The two leftmost red arrows in late-2011 are the time
    the Calar Alto low-resolution spectra were obtained, and the red
    photometric point below the second arrow is the Calar Alto
    photometric measurement. The rightmost blue arrow in early 2012
    corresponds to the time the ESO/XSHOOTER spectrum was obtained.}
              \label{lc}%
    \end{figure*}

    The long-term light curve of AA Tau in the V filter recorded over
    26 years is shown in Fig.~\ref{lc}. From 1987 till the beginning
    of 2011, the system was almost never fainter than V=14 mag,
    even in the center of eclipses. During the 20 seasons over this
    time span for which we gathered enough data, the maximum
    brightness level smoothly varied within the range 12.3-12.7~mag,
    the average brightness level remained between 12.7 and 13.1~mag,
    and the seasonal photometric amplitude varied between 0.6 and
    1.7~mag in the V band (cf. Grankin et al. 2007). Periodic light
    variations are present over this time span with a period of 8.19d
    (Artemenko et al. 2012).  However, at some point in 2011 the
    average brightness level suddenly decreased by $\simeq$2~mag,
    from V=12.8 to V=15, and the maximum brightness level dimmed
    below V=14. 

\begin{figure}
   \centering
 \includegraphics[width=\hsize]{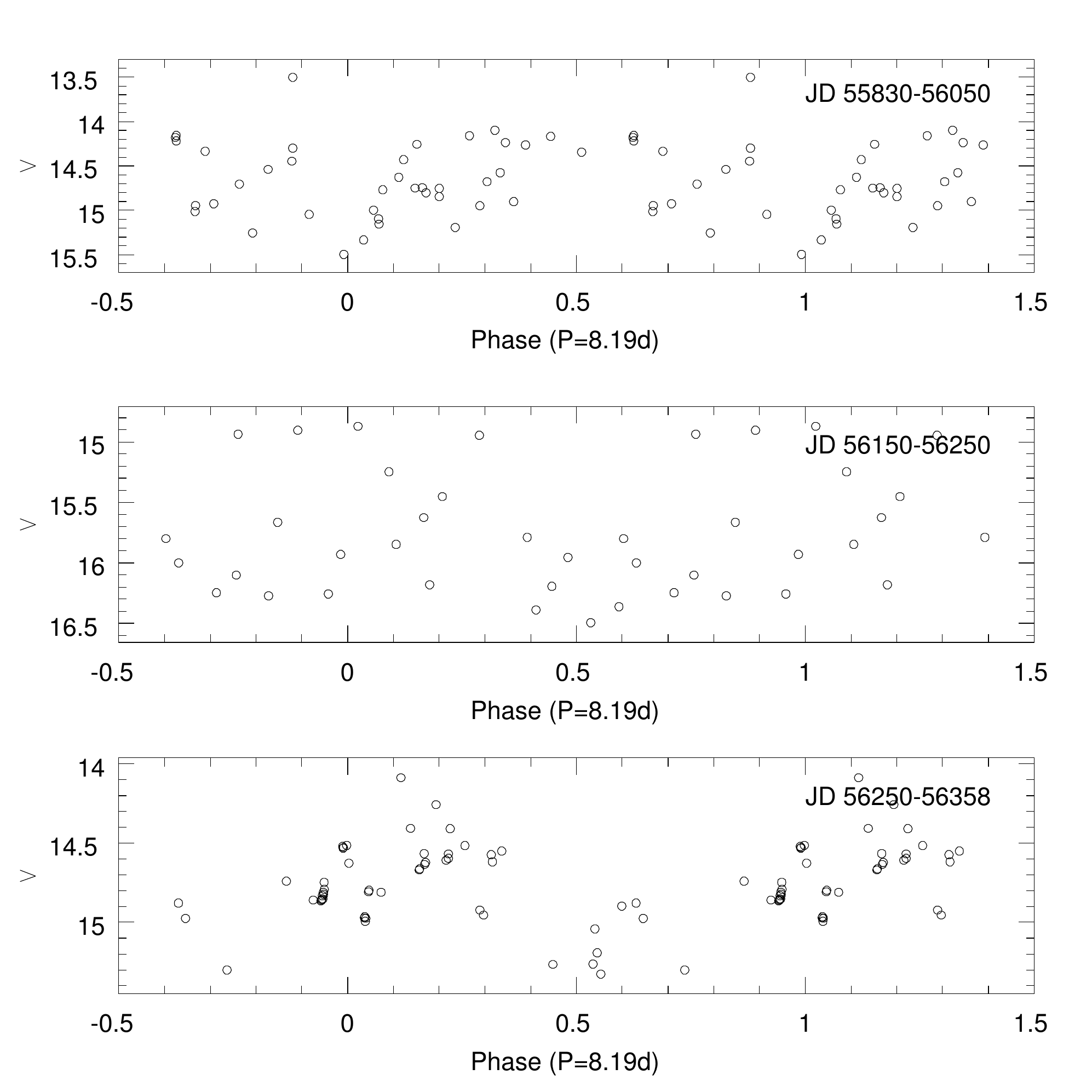}
 \caption{AA Tau's deep state light curve folded in phase assuming a
   period of 8.19d. {\it Top panel:} 2011/12 season (JD
   55830-56050). {\it Middle panel:} First half of the 2012/13 season
   (JD 56150-56250); {\it Bottom panel:} Second half of the 2012/13
   season (JD 56250-56358). The origin of the photometric phase is
   arbritrary. }
              \label{phase}%
    \end{figure}

    An enlargement of the 2011-2013 seasons is shown in
    Fig.~\ref{lc}. The deep minimum has continued for more than
    500~days. The lowest brightness level (V$\simeq$16.5~mag) was
    registered at the end of September 2012. From 20 November 2012 on
    (JD$\geq$56251), the average brightness level apparently
    remained roughly constant at V$\simeq$14.8, and the photometric
    amplitude reached up to 0.9~mag. Both mid-term change, on a
    timescale of several weeks (e.g. JD56150-JD56250), and short-term
    variations, up to 0.4-0.6~mag per day, are recorded. In
    particular, an UX~Ori-like smooth and long minimum, with little
    photometric dispersion, is quite noticeable in the fall of 2012.
    A search for periodicities in the 2011/12 and 2012/13 photometric
    datasets, analyzed either separately or combined, did not reveal
    any significant periodic signal. To investigate the
    existence of the 8.19d period that the system exhibited over more
    than 20 years in the bright state (Grankin et al. 2007; Artemenko
    et al. 2012), we nevertheless examined phase diagrams for various
    parts of the deep-state light curve. The results are shown in
    Fig~\ref{phase}. Neither the 2011/12 season nor the first half of
    the 2012/13 season reveal any sign of coherent photometric
    variations on a 8.19d timescale. However, surprisingly enough, the
    phase diagram for the second half of the 2012/13 season (JD
    56250-56358) strongly suggests that the 8.19d period is recovered
    over this specific $\sim$3 month time interval with a V-band amplitude
    of about 0.7~mag.

% Same result for the R-band. However, the (V-Rj) color appears to be
% periodic at 8.19d in the early 2012/13, and not in the ohter
% segments !!?? 

%

%{\bf Q: Interestingly, the maximum brightness level of AA Tau appears
%  to be modulated on a period of about 10 years. The deep state seem
%  to occur at the expected time for a new minimum of the maximum
%  brightness level. Is there something to understand here? [tbc, KG]}

\begin{figure}
   \centering
  \includegraphics{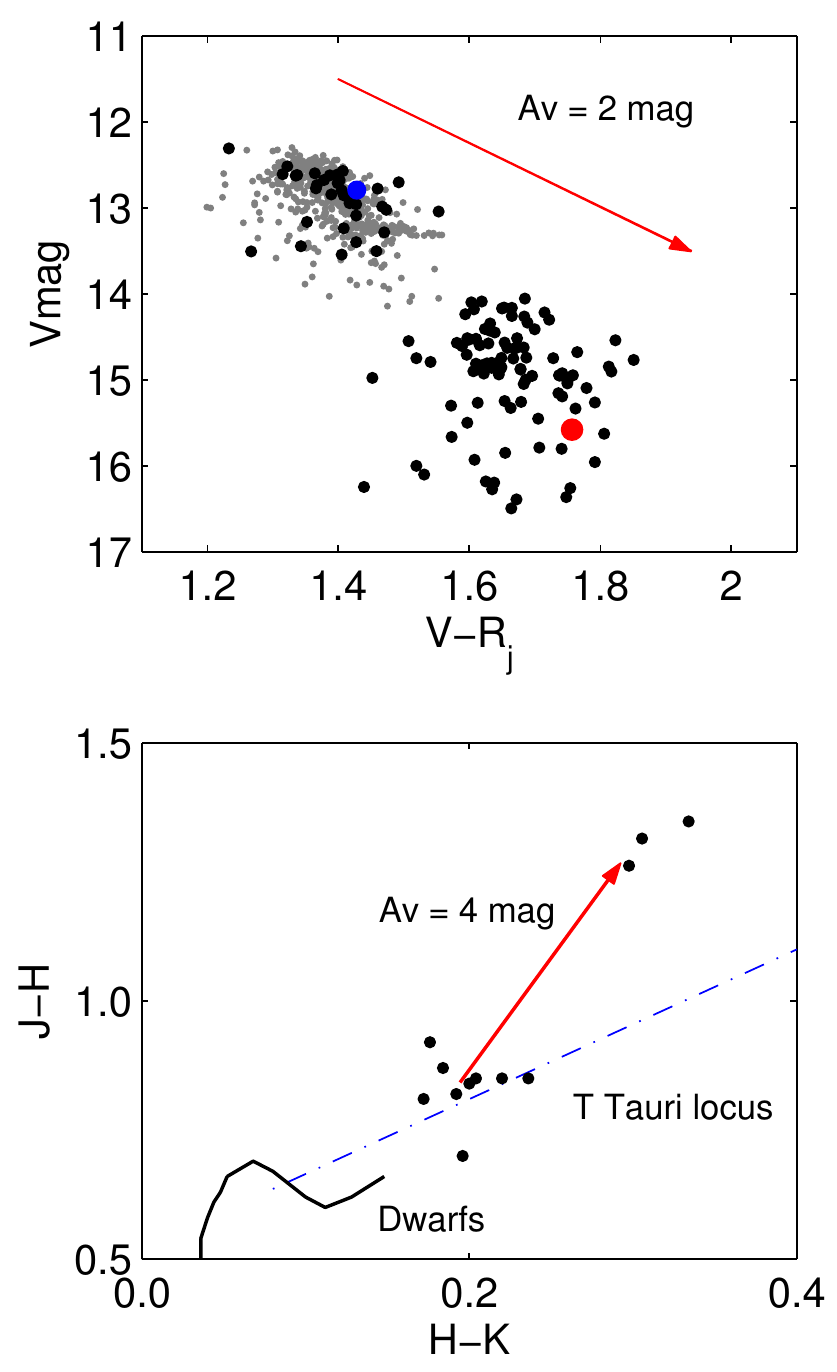}
  \caption{AA Tau's color behavior. {\it Top panel:} (V, V-R$_J$)
    color-magnitude diagram. The symbols are the same as in
    Fig.~\ref{lc}. The arrow indicates the color slope expected for
    A$_V$=2~mag of interstellar extinction. The large blue circle
    illustrates the bright state of the system, as recorded on 19 Nov.
    1995 (from Bouvier et al., 1999). The large red circle corresponds
    to the deep-state photometry obtained on 23 Dec 2011. The system
    becomes bluer at the faintest brightness level. {\it Bottom
      panel:} (J-H, H-K) color diagram. The black curve indicates the
    dwarf locus, and the blue dashed-dotted curve the T Tauri
    locus. The arrow indicates the color slope expected for
    A$_V$=4~mag of ISM-like extinction. JHK photometry for the bright
    state is taken from Bouvier et al. (1999) and for the faint state
    from Calar Alto measurements listed in Table~\ref{phot}. }
              \label{col}%
    \end{figure}

    The (V, V-R$_j$) color behavior of the system is shown in
    Fig.~\ref{col} over the full 26-year time span of AA Tau
    monitoring.  As long as the star remained brighter than V=15, the
    color variations of the system are nearly parallel to, though
    slightly bluer than, the reddening slope expected for small
    ISM-type grains. In contrast, the system does not redden more when
    it becomes fainter than V=14.5~mag. Instead, the full range of
    (V-R$_j$) colors from 1.5 to 1.8~mag is observed between V=15 and
    V=16.5~mag. This blueing effect is typical of UX Ori type
    variables (Grinin et al. 1991; Herbst et al. 1994), whose
    photometric behavior is caused by variable circumstellar
    extinction. It is usually interpreted as scattered light becoming
    a major component of the system's flux as direct light from the
    central object is depressed.

% Need to have a look at the (V-R) light curve : cf. v_vr.pdf in
% ./CrAO. The (V-R) colors seem pretty stable (and unrelated to
% brightness) in the deep state == opaque screen ? 

    Fig.~\ref{col} also shows the near-infrared (J-H, H-K) color-color
    diagram of the system as it switches from the bright to the deep
    state. In the bright state, AA Tau is located close to the T Tauri
    locus, indicative of moderate accretion and little extinction, in
    agreement with previous estimates (e.g. Bouvier et al. 1999). In
    the deep photometric state, the near-IR colors of the system are
    much redder. The transition between the bright and deep states in
    this diagram suggests a 4~mag increase of the visual extinction on
    the line of sight.  As discussed below, since the system is seen
    nearly edge-on, the sudden increase of extinction on the line of
    sight is probably of circumstellar origin.

\subsection{Spectroscopy}

% [OI] lines Hartigan et al. 2004; 1995;  Edwards
% et al. (1987)
% [OI]6300, 6364, 5577; [SII] 6731, 6716; [NII]6584

%Nov.95 (B99) Ha:10-45  Hb:2-11
%Nov.04 (B07) Ha:7-64 Hb:0.7-4 HeI:0.2-1.6 
%Jan.09 (D10) Ha:6-45 Hb=0-7 HeI:0.2-0.9

   \begin{figure}
   \centering
   \includegraphics[width=\hsize]{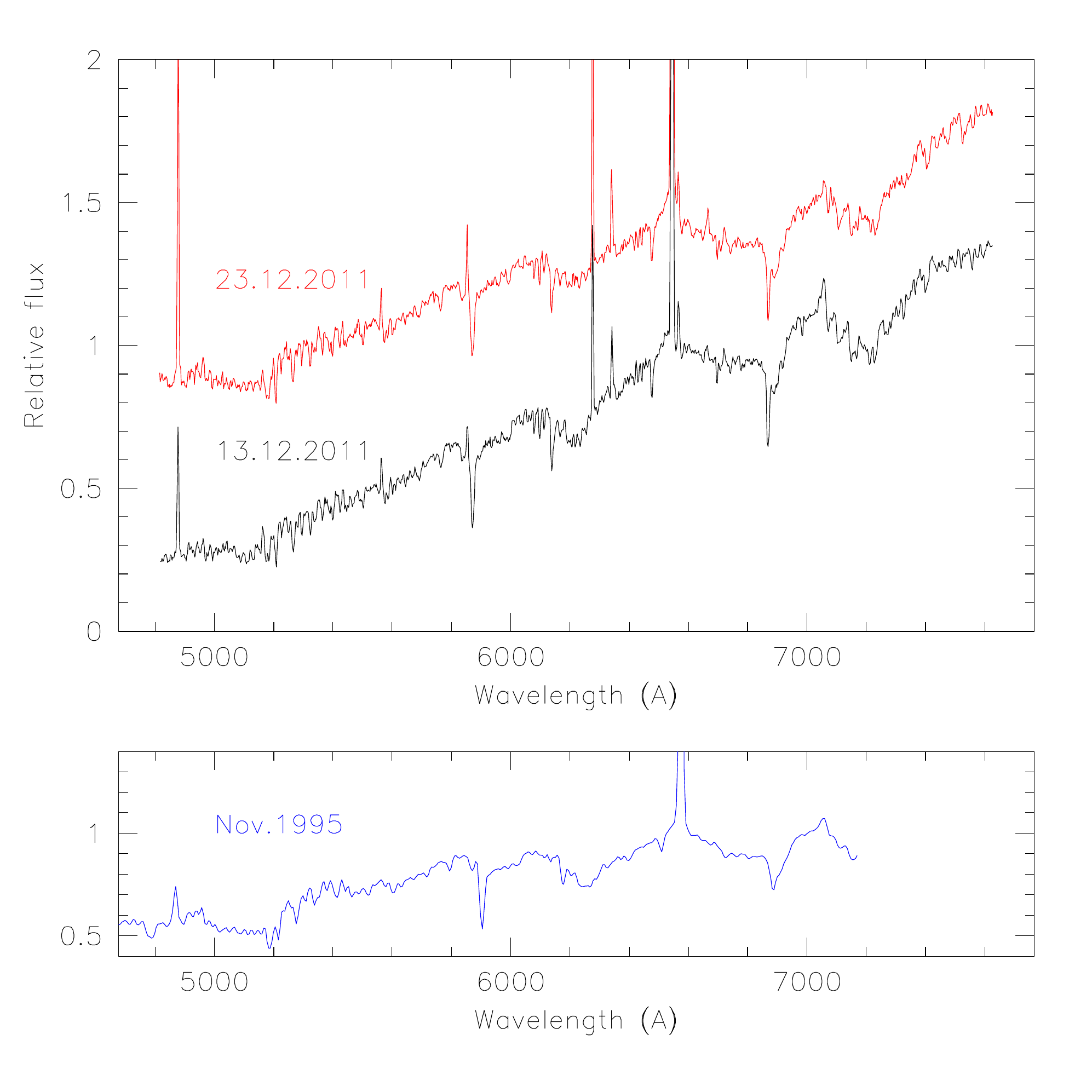}
   \caption{Low-resolution optical spectra of the AA Tau system. {\it
       Top panel:} The Calar Alto Cafos spectra obtained on Dec. 13
     and 23 Dec. 2011 are shown over the wavelength range
     480-770~nm. The continuum has been normalized to unity at the
     base of the H$\alpha$ line and the flux of the latter spectrum
     shifted by +0.5 for clarity. {\it Lower panel:} The average of 57
     AA Tau lowres spectra obtained in the bright state in Nov.~1995 is
     shown for comparison (from Bouvier et al. 1999). }
              \label{caha}%
    \end{figure}

    The low-resolution AA Tau spectra obtained on Dec. 13 and 23,
    2011, are shown in Fig.~\ref{caha} over the wavelength range
    480-770nm. A clear photospheric late-type spectrum is seen as well
    as a few emission lines including H$\alpha$, H$\beta$, HeI
    587.6~nm, and the [OI] 630.0/636.4~nm doublet. The two spectra
    obtained during the deep state are compared with the average of 57
    low-resolution spectra obtained in Nov. 1995, i.e., when AA Tau
    was still in a bright state (cf. Bouvier et al. 1999). The
    photospheric features have not changed significantly between the
    two epochs, which suggests that AA Tau has retained a spectral
    type of K7 in the deep state. A major difference, however, is the
    continuum slope, which is much steeper in the 2011 spectra than in
    those taken in the bright state, which indicates increased
    reddening. Assuming a standard interstellar extinction law
    (R$_V$=3.1, Cardelli et al. 1989), we attempted to reproduce the
    continuum slopes of the 2011 spectra by reddening the 1995 bright
    state spectrum. We obtained good fits to the 2011 Dec. 13 and 23
    continuum spectral slopes by adding 2.4~mag and 1.1~mag of visual
    extinction to the 1995 spectrum, respectively. The system had a
    V-band magnitude of V=15.58 at the time the Dec.23 spectrum was
    obtained, i.e., about 2.8~mag fainter than AA Tau's bright
    state. Yet, the spectral slope suggests an additional extinction
    amounting to only 1.1~mag compared with the bright state. This
    confirms that the system becomes bluer when it is fainter, as
    reported from the photometry above. A similar behavior was
    already observed in the changing continuum slope of the 1995 spectral
    series (cf. Bouvier et al. 1999) and was ascribed to the increasing
    contribution of scattered light as the system dims.

\begin{table}
\caption{Equivalent width (\AA) of emission lines in CAHA spectra.}             % title of Table
\label{ewcaha}      % is used to refer this table in the text
\centering                          % used for centering table
\begin{tabular}{l l l l l l l}        % centered columns (4 columns)
\hline\hline                 % inserts double horizontal lines
Line & H$\alpha$  &    H$\beta$  &   HeI &    [OI] &[OI]
& LiI  \\
$\lambda$ (nm) & 656.3 & 486.1 & 587.6 & 630.0 & 636.4 & 670.7 \\
\hline
13.12.2011   & 31.4 &   10. &   0.8   &    5.8   &      1.5   &       -0.49\\
23.12.2011   & 70.    &  19.  &  1.6    &   7.5    &      2.1   &
-0.36\\
\hline
\end{tabular}
\end{table}
  
Another clear difference between the 1995 and 2011 spectra are 
the additional emission lines in the latter. While the
H$\alpha$ and H$\beta$ lines were already observed in emission in 1995,
the new spectra additionally exhibit clear emission in the HeI 587.6
and [OI] 630.0/636.4~nm lines\footnote{We disregard the weak
  [OI]557.7~nm emission as it is significantly polluted by sky
  brightness and could partly result from unperfect sky subtraction.}.
The emission line equivalent widths were measured on both 2011 spectra
by Gaussian fitting to the line profiles. The results are listed in
Table~\ref{ewcaha}. Clearly, the strength of the Balmer and HeI
lines in the Dec.23 spectrum is twice as large as it was in the Dec.13
one. Whether this reflects actual line flux variation or merely
a variable continuum level, we cannot tell because we do not have simultaneous
photometry for the latter spectrum. However, we notice that the
equivalent width of the [OI]630.0/636.4~nm lines also increased by a
factor of about 1.4 between the two spectra. The [OI] lines are thought
to form far from the star, presumably in the large-scale jet recently
imaged by HST (Cox et al. 2013), and are therefore not expected to
intrinsically vary on such a short timescale (cf. Bouvier et
al. 1999).  This suggests that the continuum level of the Dec.23
spectrum was indeed about a factor of 1.4 lower than that of the
Dec.13 spectrum. The dimming of the continuum may thus at least partly
account for the enhanced equivalent width of the Balmer and HeI lines
in the Dec.23 spectrum, whose slope is also bluer than that of the
Dec.13 spectrum.

Combining the measured equivalent widths with the photometric
measurement obtained simultaneously with the Dec.23 spectrum, we
derived the H$_\alpha$ and H$_\beta$ emission line flux. The observed
line flux is given by $F(H_\alpha)\simeq F^0_R\cdot EW(H_\alpha) \cdot
10^{(-0.4\cdot m_R)}$ and $F(H_\beta)\simeq F^0_B\cdot EW(H_\beta) \cdot
10^{(-0.4\cdot m_B)} $, where $m_R$ and $m_B$ are the V- and B magnitudes of
the system, while $F^0_B=1880$~Jy and $F^0_R=3080$~Jy are the fluxes
for a zero magnitude star in the B- and V band, respectively. This
yields $F(H_\alpha)$=2.9 10$^{-13}$ and $F(H_\beta)$=1.1 10$^{-14}$
erg.cm$^2$.s$^{-1}$. For a distance of 140~pc, the line fluxes
translate into line luminosities of $\log L(H_\alpha)/L_\odot$=-3.74
and $\log L(H_\beta)/L_\odot$=-5.16.
% L(LHa) = 6.9e29 == 1.8 e-4 Lsun 
% $L(Hb)= 2.6e28 ==  6.8e-6 Lsun
Using the calibration between line and accretion luminosities proposed
by Fang et al. (2009) and Rigliaco et al. (2012), we derive mass
accretion rates of 2.9-4.4 10$^{-10}$ M$_\odot$yr$^{-1}$ from
H$_\alpha$ and 1.4-4.5 10$^{-11}$ M$_\odot$yr$^{-1}$ from
H$_\beta$. If the emission line regions suffer the same amount of
reddening as the stellar photosphere, these estimates have to be
corrected for extinction. Assuming that the photometric variations are
mostly driven by circumstellar extinction, and taking V$\simeq$12.8 as
the reference bright level for the system (cf. Bouvier et al. 1999),
we derive $\delta$A$_V$$\simeq$2.8~mag from the observed V=15.58 at
the time the spectrum was taken. Correcting for the enhanced
extinction, the mass accretion rate becomes $\simeq$7 10$^{-9}$ and
$\simeq$2 10$^{-9}$ M$_\odot$yr$^{-1}$ from H$_\alpha$ and H$_\beta$,
respectively. This range of mass accretion rates is similar to what
has been derived for AA Tau in the bright state (e.g. Donati et
al. 2010; Valenti et al. 1993). Clearly, the sudden dimming of the
system is not associated with a massive accretion outburst or
quenching.

%{\bf Q: the [OI] line if formed in the jet shold not be affected by
%  extinction. Has their EW varied between bright and faint state?
%Hartigan et al. (1990) EW(Ha)=26-60, EW[OI]=0.6-2.1A (r=0.39-0.77)
%Hartigan et al. (1995) EW[OI]=0.6-2.4A, [NII]=0.04-0.25A [SII]<=0.06 (r=0.15-0.63)
% factor of 6 difference between [OI] bright and [OI] faint == ~2 mag,
% like spectral slopes

%{\bf Q: any evidence for Ba II? : not int the XSHooter spectrum 
 
\begin{table*}
  \caption{Equivalent width of emission lines recorded in the ESO/XSHOOTER spectrum.}             % title of Table
\label{eweso}      % is used to refer this table in the text
\centering                          % used for centering table
\begin{tabular}{l l l | l l l | l l l }        % centered columns (4 columns)
\hline\hline                 % inserts double horizontal lines
Line & $\lambda$  & EW  & Line & $\lambda$  & EW  &
Line & $\lambda$  & EW  \\
& (nm) & \AA && (nm) & \AA && (nm) & \AA \\
\hline
 H16 &370.5 &2   &	 HeI & 447.2 &  1.2  &	 [SII] & 673.1 &  0.17 \\
 H15 &371.3 &2.5   &	 HeII & 468.7 &  0.4  &	 CaII & 849.9 &  0.54 \\
 H14 &372.2 &3.0   &	 H$\beta$ & 486.1 &  9.0   &	 CaII & 854.3 &  0.72  \\
 H13 &373.5 &4   &	 [OI] & 557.8 &  0.6  &	 CaII & 866.3 &  0.46 \\
 H12 &375.1 &7   &	 [SII] & 407.7 &  0.8  &	 Pa$\delta$ & 1004.9 &  1.4  \\
 H11 &377.1 &11   &	[SII] & 406.86 &  4.0  &	 HeI & 1083. (em.) &  2.0  \\
 H10 &379.8 &15   &	HeI & 587.6 &  1.2   &	HeI & 1083. (abs.) &  2.5 \\
 H9 &383.6 &20  &	 [OI] & 630.0 &  3.8   &	 Pa$\gamma$ & 1093. & 1.4  \\
 H8 &388.9 &26   &	 [OI] & 636.4 &  0.9   &	Pa$\beta$ & 1282. & 1.5 \\
 CaIIK &393.4 &14.5   &	 H$\alpha$ &  656.3 & 26.0   &	 H$_2$ & 2122. & 0.3 \\
 CaIIH/H$\epsilon$ &397.0 &31   &	 [NII] & 658.3 &  0.3  &	 Br$\gamma$ & 2166. &  no emission \\
 H$\delta$ & 410.2 &13   &	 HeI & 667.9 &  0.3 \\	
 H$\gamma$ & 434.1 &  14   &	 LiI & 670.7 & -0.5  \\	
\hline
\end{tabular}
\end{table*}

   \begin{figure}
   \centering
   \includegraphics[width=\hsize]{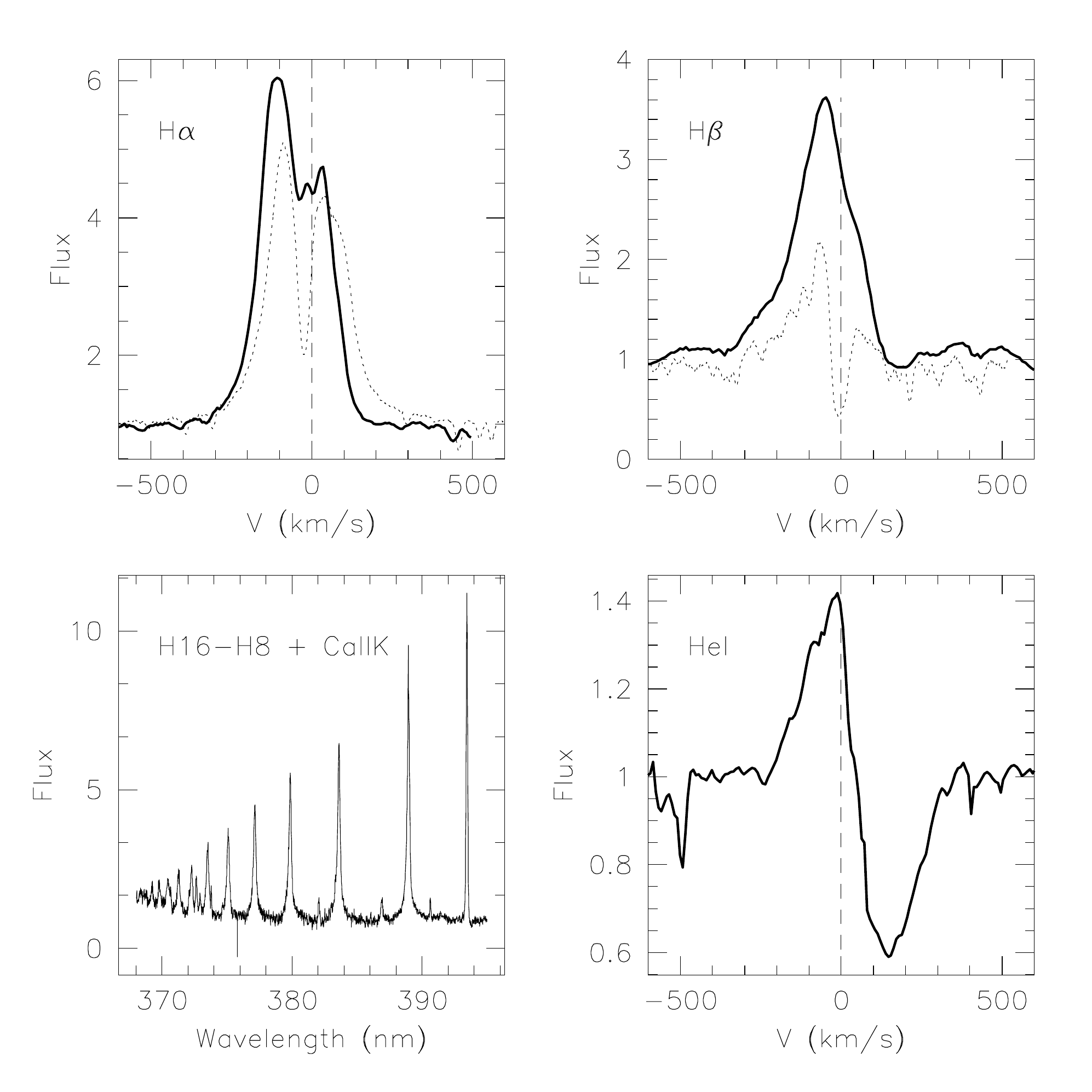}
   \caption{Line profiles from the mid-resolution VLT/XSHOOTER
     spectrum obtained on Feb.25, 2012. {\it From top to bottom and
       left to right:} H$_\alpha$, H$_\beta$, the higher Balmer series
     and Ca~IIK, and HeI 1083~nm. The lower axis is labeled with
     velocities corrected for the star's radial velocity
     (V$_{rad}$=17.2~km~s$^{-1}$; Donati et al. 2010), except for the
     higher Balmer series where it is labeled in wavelength. The
     continuum flux has been scaled to unity for all profiles.
       The dotted H$_\alpha$ and H$_\beta$ profiles shown in the upper
       panels are the mean profiles obtained during AA Tau's bright
       state by Bouvier et al. (2007).}
              \label{prof}%
    \end{figure}
    The higher resolution XSHOOTER spectrum offers a deeper insight
    into the shape of the emission line profiles, a few of which are
    shown in Fig.~\ref{prof}. The H$_\alpha$ and H$_\beta$ line
    profiles obtained in the deep state are clearly asymmetric and
    purely in emission. Most noticeably, they lack the deep central
    absorption seen in similar profiles obtained during the bright
    state (cf., e.g., Bouvier et al. 2003, Fig.8; Bouvier et al. 2007,
    Fig.8; Donati et al. 2010, Fig.3). The HeI 1083~nm line profile
    has a clear inverse P Cygni signature with a deep redshifted
    absorption indicative of accretion onto the central object. The
    profile is qualitatively similar to those reported by Fischer et
    al. (2008) and obtained in 2002-2006 while AA Tau was still in the
    bright state. The equivalent width of the HeI redshifted
    absorption component we measure on the XSHOOTER spectrum is
    2.6\AA, directly in the center of the range measured for this feature
    by Fisher et al. (2008). Moreover, the redshifted velocity at which
    the absorption component reaches back to the continuum is
    $\sim$320~km.s$^{-1}$, similar to that reported in bright state
    HeI line profiles. Many other lines are seen in emission in the
    XSHOOTER spectrum, and their equivalent widths are listed in
    Table~\ref{eweso}.

% NaD shows 2 narrow components : interstellar or circumstellar ? 
% The first is at 589.6925/589.692 nm, the second at 589.0995/589.096
% nm; intrinsic lambda NaD1=589.592 nm; NaD2=588.995 nm. Yields
% Vrad=51 km/s and 51km/s, respectively.  
% LiI is at 670.88nm; intrinsic lambda is 6707.76+6707.91
% CaI is at 671.8654/671.865nm; intrinsic is 671.769nm; yields 43 km/s
% Is the VIS spectra corrected for HELICOR/BARYCOR=-30.13 km/s ??? No,
% it is not. Then, all is in place. Also, i checked the 2004 HARPS NaD
% profiles, they are the same as the xsh ones. Nothing new there
% (although the narrow --circumstellar?- components EW is reduced
% by a factor~2 compared to 2004). 

    Although we do not have photometric measurements simultaneous to
    the XSHOOTER spectrum obtained on 25 Feb. 2012 (JD 55983.0270), V-
    and R-band measurements were taken a few nights apart, on 23 and
    26 Feb. 2012. Averaging these measurements and combining them to
    EW($H_\alpha$)=26\AA, we derive from the extinction-corrected
    H$_\alpha$ line luminosity a mass accretion rate of 3.6
    10$^{-9}$~M$_\odot$yr$^{-1}$, in agreement with that derived above
    from the Calar Alto spectrum obtained two months earlier. Hence, we
    do not find evidence for significant variations of the mass
    accretion rate on this timescale during the deep state. 

\begin{figure}
   \centering
   \includegraphics[width=\hsize]{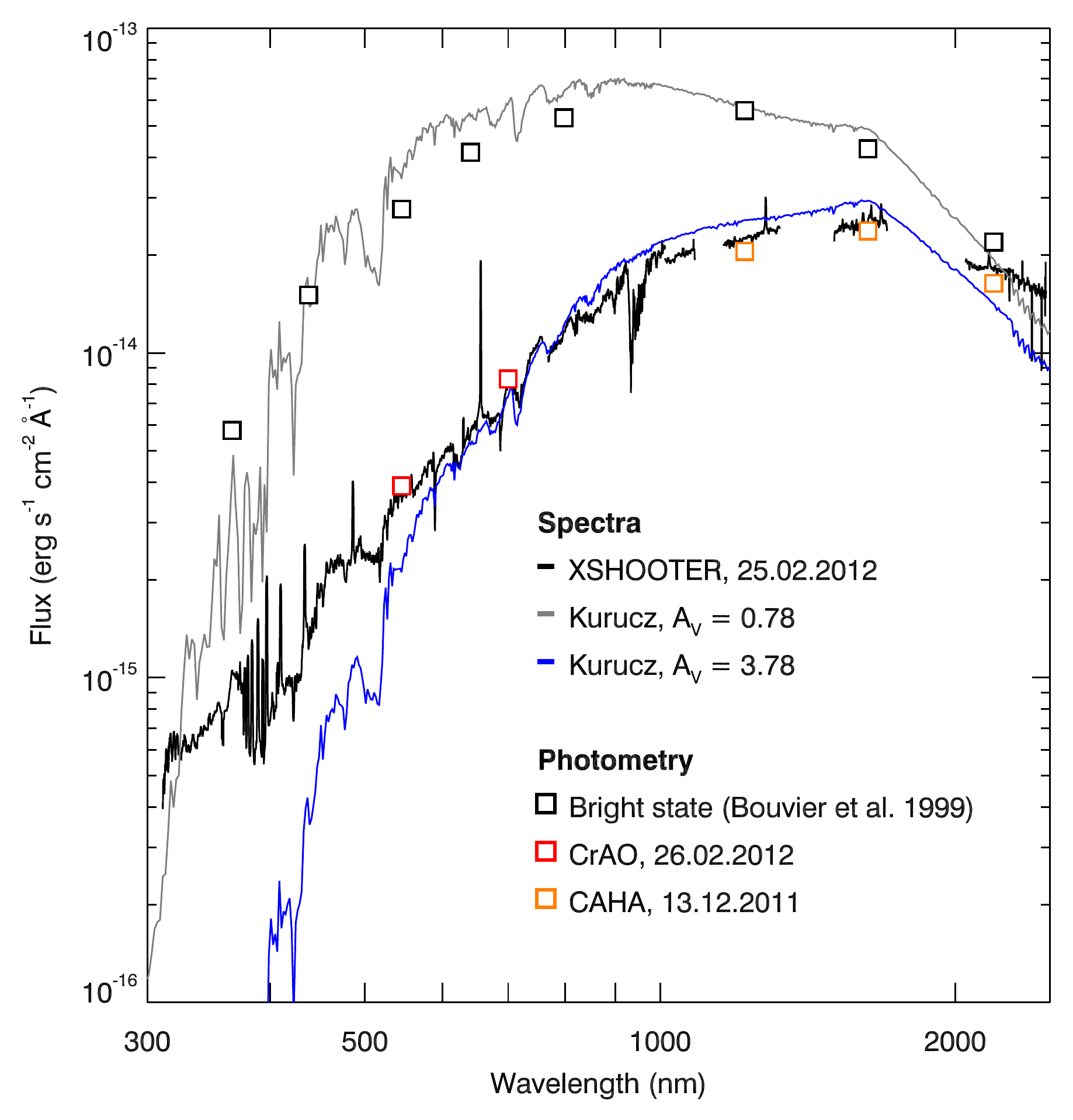}
  \caption{Flux calibrated VLT/XSHOOTER spectrum obtained on
     Feb.~25, 2012, is shown (thick black curve) over the 310-2470~nm
     wavelength range on a logarithmic scale. Regions affected by
     strong telluric absorption in the JHK bands are clipped from the
     spectrum. The red squares corresponds to V and R$_J$ photometry
     obtained at CrAO on Feb.~26, 2012 (V=14.92, R$_J$=13.30), and the
     orange squares to JHK photometry obtained at Calar Alto on
     Dec.13, 2011. The upper spectrum (gray curve) is a Kurucz model
     computed for AA Tau's parameters reported in Bouvier et
     al. (1999) (T$_{eff}$=4000~K, R=1.85~R$_\odot$, d=140~pc,
     A$_V$=0.78~mag) and the overplotted black squares correspond to
     the UBVRIJHK bright-state photometry from the same study. The
     lower spectrum (blue curve) is the same Kurucz model reddened by
     A$_V$=3.78~mag.}
              \label{xsh}%
    \end{figure}

    The full XSHOOTER spectrum is shown in Figure~\ref{xsh} and
    is compared with Kurucz models computed for AA Tau's parameters
    (T$_{eff}$=4000~K, R$_*$=1.85~R$_\odot$, d=140~pc; from Bouvier et
    al. 1999). A slightly reddened Kurucz spectrum, with A$_V$=0.78~mag,
    provides a good match to the bright-state UBVRIJHK photometry of
    the system. A much higher extinction, A$_V$$\sim$3.78~mag, is
    required to match the red and near-IR regions of the XSHOOTER
    spectrum obtained in the deep state. This result supports a
    $\geq$3~mag increase of visual extinction on the line of sight as
    being the cause of AA Tau's sudden deepening. At wavelengths
    shorter than 700~nm, the system's flux is higher than predicted by
    the reddened Kurucz model, suggesting that the system's luminosity
    at shorter wavelengths is dominated by scattered light.

\section{Discussion}
\label{dis}

Since it began to be monitored in 1987 at Mount Maidanak Observatory
and until 2011, AA Tau has exhibited a relatively stable photometric
behavior. Yearly light curves exhibit a nearly constant maximum
brightness level (V$\simeq$12.3-12.7) interrupted by periodic dips
with an amplitude up to 1.5~mag occurring every 8.2~days and lasting
for a few days (e.g. Artemenko et al. 2012). The 8.2d period had
   originally been reported by Rydgren et al. (1985) from photometry
  obtained in 1984. The dips have been interpreted has eclipses of the
central star by a rotating inner disk warp located at a distance of
8.8R$_*$ from the stellar surface (Bouvier et al. 1999; M\'enard et
al. 2003). The nonaxisymmetric inner disk warp was ascribed to the
interaction between the inner disk and an inclined stellar dipole
(Bouvier et al. 1999, 2003, 2007). Follow-up spectropolarimetric
observations confirmed the strong (2-3~kG) and slightly
tilted ($\sim$20$\degr$) stellar dipole (Donati et al. 2010).
%Incidentally, the 20yr-long light curve shown in
%Fig~\ref{lc} suggests that the maximum brightness level may undergo a
%low amplitude modulation on a timescale of about 10 years, a point to
%which we return below.

At some point between spring and fall 2011, AA Tau suddenly faded by
$\sim$2~mag in the V band and remained at around this deep photometric
state since then. By comparing the CCD image taken at Calar Alto
  on Dec.~2011 with the SDSS9 catalog release (Ahn et al. 2012), we
  verified that the three field stars nearest to AA Tau, located at a
  projected distance ranging from 26 to 53$\arcsec$, did not exhibit
  any luminosity variation, thus confirming that the sudden dimming is
  indeed restricted to AA Tau.  We additionally constrained the spatial
  extent of the source of enhanced extinction by investigating whether
  AA Tau's jet has undergone a similar deepening as the central
  star. Assuming that the [OI]~630.0~nm line forms in the jet, we
  computed the line's luminosity for both the bright and the deep
  photometric states, as $F([OI])\simeq F^0_R\cdot EW([OI]) \cdot
  10^{(-0.4\cdot m_R)}$. For the bright state, we combined the
  measurement from Hartigan et al. (1995) EW([OI])=0.58\AA\ obtained
  on JD~2447169.8 (Jan.1988) with Maidanak's photometric measurement
  $R$=11.46~mag secured on JD~2447170.4, to obtain
  $F([OI])$=3.2~10$^{-14}$~erg.cm$^2$.s$^{-1}$. For the deep state, we
  used our Calar Alto measurement from Dec.~23, 2011, EW([OI])=7.5\AA\
  and the simultaneous photometric measurement $R$=14.35, which yields
  $F([OI])$=2.7~10$^{-14}$~erg.cm$^2$.s$^{-1}$. We conclude that the
  jet's luminosity, as measured by the [OI] line flux, has not changed
  between the bright and deep states. The source of AA Tau's deepening
  event must therefore be more compact than the jet, at least along
  the jet direction, which extends to about 21$\arcsec$ at 
  P.A.=183$\degr$ from the central star (Cox et al. 2013).

The results presented above and our previous
knowledge of the system prompt us to interpret this fading as the
consequence of enhanced extinction on the line of sight produced by
dust in the disk. Both the near-IR (J-H, H-K) colors of the system in
the deep state and the slope of the red and near-IR parts of the
XSHOOTER spectrum suggest that the system is suffering an additional
3-4~mag of visual extinction compared with the bright state
(A$_V$=0.8~mag, Bouvier et al. 1999). The V-band magnitude difference
between the bright and deep state is, however, only of about 
2-3~mag, and the slope of the optical spectra is also bluer than
expected for a 4~mag increase of visual extinction. Indeed, the system
becomes bluer at optical wavelengths as it reaches minimum brightness
(see Fig.~\ref{col}), a color behavior typical of UX Ori-type
variables, whose variations result from obscuration by circumstellar
dust (Grinin et al. 1991; Herbst et al. 1994). Altogether, this
suggests that, as direct photons are being absorbed by the dusty disk
on the line of sight, scattered light largely contributes to, and even
dominates the system's optical flux at minimum brightness.

This interpretation provides a consistent framework for
understanding not only the photometric properties of the system in the
deep state and its spectral slope from UV to near-IR wavelengths, but
also accounts for the disappearance of the 8.2d periodic eclipses from
the optical light curve as well as for the lack of central absorption
components in the Balmer line profiles, two features regularly
observed in the bright state. Assuming scattered light is a major
component of the system's optical flux in the deep state, while direct
light is largely suppressed, the photons reaching the observer must
have escaped the central system toward unobscured lines of sight,
i.e., polewards or at least toward relatively high latitudes. At
these angles, reaching well above the disk midplane, the stellar
photons do not suffer occultations by the inner disk warp, and
therefore do not convey any periodic photometric signal as they are
scattered back to the observer. The late recovery of the 8.19d period at
the end of the 2012/13 season suggests that the scattered/direct light
fraction has changed by then.
%possibly by the surface layers of the disk (cf. Cox et al. 2013).  
Similarily, the deep central absorption component of the Balmer line
profiles, always present in the bright state, presumably arises from
the photons being absorbed either very close to the star by the
slow-moving material at the top of the accretion funnel flows or
farther away at the base of a dense disk wind nearly perpendicular to
the line of sight (Kurosawa et al. 2006, 2011). Scattered photons that
have escaped the central system along trajectories located well above
the disk midplane do not encounter either of these absorbing media and
therefore do not exhibit central absorption signatures in the line
profiles (cf. Appenzeller et al. 2005; Grinin et al. 2012). Only the
HeI 1083~nm line still displays a pronounced redshifted absorption
component in the deep state that is essentially unchanged from the
bright state, which is indicative of accretion onto the central star
(Fischer et al. 2008). Indeed, at this wavelength scattered light does
not dominate any more (cf. Figs.~\ref{col} and \ref{xsh}) and the
central part of the system, though heavily reddened, can be directly
seen.

Because AA Tau is surrounded by a circumstellar disk seen nearly
edge-on (Cox et al. 2013), the deep photometric state the system
currently undergoes most likely results from a drastic change in the disk
geometry, at least on the line of sight. What is the physical process
at the origin of the enhanced extinction on the line of sight, and at
which distance in the disk is the obscuring material located?

Star-disk interaction MHD models indicate that various kinds of
instabilities may occur at the inner disk edge. For instance, Romanova
et al. (2013) have shown that one-arm bending waves can develop as the
inner disk responds to an inclined stellar dipole, which results in an
inner disk warp whose scale height $h$ may exceed 0.3$r$. Such a warp
is held responsible for AA Tau's eclipses in the bright state (Bouvier
et al. 1999) and an average warp scale-height of $h/r$$\sim$0.3 has
been deduced from the statistical analysis of an ensemble of AA
Tau-like light curves in NGC2264 (Alencar et al. 2010). The warp
geometry is sensitive to the topology and inclination of the stellar
magnetic field and to the mass accretion rate in the inner disk
(Romanova et al. 2012). While the deepening of the AA Tau system could
be tentatively ascribed to a sudden increase of the warp scale height,
resulting in the total occultation of the central star, we consider
this interpretation unlikely. Firstly, any change in the inner disk
property would probably affect the mass accretion rate onto the
star. Yet, our estimates of mass accretion rate obtained in the deep
state are not significantly different from those measured in the
bright state. Second, the near-IR colors of the system in the deep
state are reddened by at least 4~mag of visual extinction. The near-IR
flux comes from the inner disk regions, which suggests that the
obscuring material is located beyond a few 0.1~AU from the central
star, i.e., much farther away than the inner disk edge
($r\sim$0.08~AU; Bouvier et al 1999). Third, AA Tau's light curve in
the last season shows a slowly varying extinction pattern with little
photometric dispersion that occurs on a timescale of
$\sim$100~days. If this timescale is to be related to Keplerian
rotation in the disk, it points to radial distances well beyond the
inner edge. More generally, the extended duration of the deepening
episode ($\geq$2yr) seems to exclude any short-timescale instability
operating in the inner disk, which would be expected to produce
transient variations of the disk scale height (e.g., Turner et
al. 2010).

Several lines of evidence thus point to the obscuring material being
located far beyond the inner disk edge. The sudden appearance of
additional extinction on the line of sight could conceivably result
from the growth of an instability that acts to increase the disk
scale-height on long timescales (e.g. Simon et al. 2011; Flock et
al. 2012; Lesur et al. 2013). This process remains speculative because 
detailed physical conditions in the disk, such as viscosity, magnetic
field, etc., are unknown. However, we note that these unstabilities may
eventually trigger FU Ori outbursts as they propagate inward (Zhu et
al. 2010; Bae et al. 2013), thus potentially making AA Tau a pre-FUOr
candidate. Disk instabilities may also trigger dense disk winds (e.g.,
Lesur et al. 2013; Bai \& Stone 2013) that may be the source of
enhanced line-of-sight extinction if they are able to lift up dust
from the disk mid-plane (Owen et al. 2011; Bans \& K\"onigl
2012). However, AA Tau's HeI 1083~nm line profile does not bear
evidence for significant wind absorption. Alternatively, enhanced
extinction could also result from an azimuthal disk asymmetry coming
onto the line of sight at Keplerian velocity. Nonaxisymmetric density
and associated scale-height bumps may be triggered in the disk by an
embedded planetary mass companion (e.g., Uribe et al. 2011; Kley \&
Nelson 2012; Baruteau \& Masset 2013) or arise from instability driven
density vortices (e.g., Lesur \& Papaloizou 2010; Meheut et al. 2012;
Lin 2012).
% another possibility is precession of a disk warped by a stellar
% companion (cf. Fragner & Nerlon 2010). 
Assuming the disk overdensity rotates at Keplerian velocity, the
observed timescales for AA Tau's photometric variations provide some
constraints on its radial distance from the central star. Because the bright
state prior to the deepening episode lasted for at least 24 years,
the orbital period of the disk perturbation must be at least this
long. Assuming Keplerian rotation around the central 0.8~M$_\odot$
star, this locates the density bump at a distance of 7.7~AU or more
from the central star. At this distance, the 2~yr duration of the
on-going deep state implies an azimuthal extent of at least 30 degrees
in the disk. Shorter timescale photometric variations observed within
the deep state, ranging from days to months (cf. Fig.~\ref{lc}),
furthermore suggest that the azimuthal distribution of dust in the density
perturbation is not homogeneous.
% note:  t_transit = 3 days

To investigate whether AA Tau could have been in such a
  faint state in previous times, we searched all available
  publications providing AA Tau's photometry up to the mid-80s, i.e.,
  before the yearly Maidanak monitoring project started, based on the
  bibliography linked to the object at CDS Strasbourg. Joy (1949)
  first reported V$\simeq$14, with the warning that ``magnitude
  estimates are rough aproximations'' because they were derived from
  low-resolution photographic spectra. The oldest actual photometric
  measurement we found was reported by Varsavsky (1960), who listed
  V=12.82 obtained in the fall of 1959. A few other measurements from
  the early 60s are listed in Smak (1964), who reported V=11.9-12.9 for
  three observations obtained in 1962, suggesting that the system was
  already in a bright state at that epoch. Nurmanova (1983) reported 20
  V-band observations of the system spanning from 1974 to 1979,
  yielding V=12.5-14.2~mag. Indeed, all published V-band values we
  could find in the literature from the 60s to the mid-80s lie in
  the range from about 12.4 to 14.2~mag, i.e., similar to the
  photometric range of the long-term Maidanak light-curve covering
  from 1987 onward. A large number of fainter measurements are
  reported in the literature, but these correspond to photographic
  magnitudes, i.e., closer to the B band. We thus did not find any
  compelling evidence for a prior deepening episode of the system
  since the late 50s. However, we cannot totally exclude it either,
  since until the yearly monitoring started in 1987 at Maidanak
  Observatory the system has been only sporadically observed and the
   duration of the deepening events is unknown as yet.

\section{Conclusion}
\label{con}

While the young, low-mass AA Tau accreting system has exhibited a
stable photometric behavior during a period extending over 24 years
(1987-2011), we have reported here its sudden dimming by about
2~magnitudes that started in 2011 and is still on-going. The
photometric and spectral properties we derived for the system in the
deep state indicate that it currently suffers $\geq$4~mag of
visual extinction on the line of sight and is dominated by scattered
light at optical wavelengths. The mass accretion rate onto the star
and in the inner disk has not significantly changed between the
previous 24~yr-long bright state and the recent dimming episode. We
proposed that the additional amount of extinction on the line of sight
is produced by a density perturbation in the disk located at a
distance of $\geq$7.7~AU from the central star. This density
perturbation might result from a low mass companion embedded in the disk
that orbits the central star at Keplerian velocity, possibly a
planetary embryo.

\begin{acknowledgements}
  It is a pleasure to thank Paula Teixeira for help in preparing the
  XSHOOTER observations, Svetlana Artemenko for assistance during CrAO
  observations, Geoffroy Lesur and Christophe Pinte for discussions
  about disk instabilities and radiative transfer, respectively. We
  thank the referee for a prompt and useful report.  This research
    has made use of the SIMBAD database, operated at CDS, Strasbourg,
    France. We acknowledge partial funding from CNRS DERCI on a
    French-Ukrainian collaborative program, as well as funding from
    Universit\'e Joseph Fourier Grenoble 1 for K. Grankin’s stay at
    IPAG. This study was supported in part by the grant ANR 2011 Blanc
    SIMI5-6 020 01 "Toupies: Towards understanding the spin evolution
    of stars" (\url{http://ipag.osug.fr/Anr_Toupies/}). We acknowledge
    financial support from CNRS-INSU's Programme National de Physique
    Stellaire.  
\end{acknowledgements}

\end{document}